\documentclass[letterpaper, 10 pt, conference]{ieeeconf}  

\IEEEoverridecommandlockouts                              

\usepackage{graphics} 
\usepackage{epstopdf}
\usepackage{epsfig} 
\usepackage{mathptmx} 
\usepackage{amsmath} 
\usepackage{amssymb}  
\usepackage{color}

\title{\LARGE \bf
Model-based versus model-free control designs \\ for improving  microalgae growth in a closed photobioreactor: \\ Some preliminary comparisons
}

\author{Sihem Tebbani$^{1}$, Mariana Titica$^{2}$,  C\'{e}dric Join$^{3, 5, 6}$, Michel Fliess$^{4, 5}$, Didier Dumur$^{1}$ 
\thanks{$^{1}$Laboratoire des Signaux et Syst\`emes (L2S), CentraleSup\'elec-CNRS-Univ. Paris-Sud, Universit\'e Paris-Saclay, D\'{e}partement Automatique, 3 rue Joliot-Curie, 91192 Gif sur Yvette, France. {\tt \small \{sihem.tebbani, didier.dumur\}@centralesupelec.fr } } 
\thanks{$^{2}$GEPEA (CNRS, UMR 6144), Universit\'e de Nantes, 37 boulevard de l'Universit\'e, CRTT-BP 406,
	44602 Saint-Nazaire, France. \newline {\tt \small mariana.titica@univ-nantes.fr} }
\thanks{$^3$CRAN (CNRS, UMR 7039)), Universit\'{e} de Lorraine, BP 239, 54506 Vand{\oe}uvre-l\`{e}s-Nancy, France. \newline
{\tt\small cedric.join@univ-lorraine.fr}}
\thanks{$^4$LIX (CNRS, UMR 7161), \'Ecole polytechnique, 91128 Palaiseau, France. {\tt \small Michel.Fliess@polytechnique.edu } } 
\thanks{$^{5}$AL.I.E.N. (ALg\`{e}bre pour Identification \& Estimation Num\'{e}riques), 24-30 rue Lionnois, BP 60120, 54003 Nancy, France. \newline
        {\tt \small \{michel.fliess, cedric.join\}@alien-sas.fr}}  
\thanks{$^{6}$Projet NON-A, INRIA Lille -- Nord-Europe, France. }    
}

\begin{document}

\maketitle
\thispagestyle{empty}
\pagestyle{empty}

\begin{abstract}
Controlling microalgae cultivation, \textit{i.e.}, a crucial industrial topic today, is a challenging task since the corresponding modeling is complex, highly uncertain and time-varying.  A model-free control setting is therefore introduced in order to ensure a high growth of microalgae in a continuous closed photobioreactor. Computer simulations are displayed in order to compare this design to an input-output feedback linearizing control strategy, which is widely used in the academic literature on photobioreactors. They assess the superiority of the model-free standpoint both in terms of performances and implementation simplicity.

\end{abstract}


\section{Introduction}
The production and the culture of microalgae play a growing industrial and commercial role (see, \textit{e.g.}, \cite{kim,spo}, and the references therein). Their relationship with renewable energy and sustainable development should also be emphasized (see, \textit{e.g.}, \cite{bib:Chevalier}--\cite{iste}).
 The corresponding cultivation systems, which are called \emph{photobioreactors}, or \emph{PBR}, give rise to challenging control questions which have already attracted a lot of attention. Most of the existing academic publications are model-based (see, \textit{e.g.}, \cite{ bib:Abdollahi}--\cite{bib:Dochain}). 
Among the various control techniques, which are often nonlinear, optimal control \cite{and}, predictive control (\cite{iste,bib:Abdollahi,paw}), adaptive control (\cite{mail,rod}),  feedback linearization (\cite{ifrim,bib:tebbani2015,bib:Dochain}), and the use of partial differential equations \cite{pde} are perhaps the most popular ones. 
Although those papers are quite promising, they suffer from the great difficulty of deriving a ``good'' mathematical modeling of the bioprocess (see, \textit{e.g.}, \cite{bib:Bernard}). It is due to
\begin{enumerate}
\item its inherent complexity,  
\item its uncertain and time-varying characteristics since a life process has to be taken into account.  
\end{enumerate}

This communication is introducing therefore a new \emph{model-free control} setting \cite{csm}, which is moreover rather easy to implement both from software \cite{csm} and hardware \cite{nice} standpoints. 
There are many concrete applications, including some patents. We select here for obvious reasons publications that are related to biotechnology: \cite{toulon}--\cite{bara}.

The performances of this approach are compared to those achieved by an input-output (I/O) feedback linearization, which is among the most widely used control design in the academic literature on bioreactors \cite{bib:Dochain}. The cultivation of the microalgae \textit{Chlamydomonas reinhardtii} is considered here. The biomass concentration will be regulated to a target value, determined so that a high level of biomass productivity is achieved. The influence of light fluctuation on the reference tracking is also taken into account.

Our paper is organized as follows. The system and its modeling are presented in Section \ref{sect.model}. Section \ref{sect.control} displays two control strategies: an I/O feedback linearizing control and a model-free one. These strategies include two steps: (i) the choice of the setpoint that leads to a high biomass productivity, (ii) the regulation of the system around this setpoint. Numerical results are provided and discussed in Section \ref{sect.simulation}. Conclusions and perspectives are developed in Section \ref{sect.conclusion}.  

\section{System description and modeling}
\label{sect.model}
In the continuous operation of PBR, the reactor is continuously fed with the liquid medium culture with nutrients; the rate of outflow is equal to the rate of inflow ($F$) and the culture volume ($V$) remains constant. The manipulated variable in this case is the dilution rate ($D=F/V$). The microorganism population grows in the medium consuming nutrients (dissolved CO$_2$, nitrogen, phosphorus). In autotrophic conditions, the solely carbon source is CO$_2$, which is provided continuously by air enriched CO$_2$ bubbled in the liquid medium; its injection depends on the pH, which is maintained at the growth optimum value \cite{ifrim}. By this way, all liquid nutrients are provided in sufficient quantities for avoiding mineral limitations. The main factor governing the growth is the light which is the energy source for the growth. PBR surface is lighted artificially or naturally. For a given PBR geometry, light intensity distribution in the cultivation medium depends on biomass concentration as well as on optical properties of the microalgae, which are determined by their shape and pigment content. In continuous, mineral nonlimiting cultivation conditions, the only limiting factor is the photon flux density received by the culture. Improving the light availability is thus a crucial aspect of biomass growth and process productivity. A low amount of light would decrease the growth. This is due to lack of energy necessary to fixate carbone. Excessive irradiance on the other hand would induce inhibition phenomena. 


%

Dynamic models describing the behavior of microalgae cultures are usually a set of nonlinear ordinary differential equations. They are mainly deduced from mass balance considerations on both liquid and gaseous phases \cite{bib:Ifrim2014}. In an ``optimal'' system ensuring nonlimiting conditions with respect to the liquid nutrients and environmental conditions, \textit{i.e.}, temperature and pH, the rate of photosynthesis and productivity are determined by the light availability \cite{bib:Takache2012}. In this case, the model is represented by one differential equation expressing biomass concentration dynamics \eqref{Eq.dX}, coupled with algebraic equations giving light profile into the culture bulk (radiative model) and kinetic law yielding local photosynthetic responses by expressing growth rate as a function of local irradiance in the bulk depth. Different kinetic models are presented in the literature \cite{bib:Bernard}. In this paper, a predictive model from \cite{bib:Takache2012} has been used as benchmark for our controllers.

The mass balance model of a continuous well-stirred PBR for biomass concentration $X$, in kg/m$^3$, is as follows: 

\begin{equation}
\label{Eq.dX}
\frac{{d X}}{{d t}} = {r_X} - D X 
\end{equation}
where $r_X$ is the growth rate (kg/m$^3$/h) and $D$ the dilution rate (h$^{-1}$). 

The kinetic model proposed by \cite{bib:Takache2012} predicts photosynthesis and respiration of microalgae from an energetic analysis.

 The biomass growth rate $r_X$  is stoichiometrically linked to the net oxygen evolution rate $\left\langle J_{O_2}\right\rangle $:

\begin{equation}
\label{Eq.rX}
r_X=\dfrac{\left\langle J_{O_2}\right\rangle M_x X}{\nu_{O_2-X}}
\end{equation}
where $M_x$ is the C-molar mass for the biomass (g mol$^{-1}$) and $\nu_{O_2-X}$ the stoichiometric coefficient.
$\left\langle J_{O_2}\right\rangle $ is calculated with:
\begin{equation}
\label{Eq.J02}
\left\langle J_{O_2}\right\rangle = \frac{1}{L}\int_{0}^{L}J_{O_2}(z) dz
\end{equation}
where $L$ is the total reactor depth, $z$ is the depth of the reactor in rectangular coordinates (since the PBR here is rectangular. The PBR considered here was presented in detail in \cite{ifrim}), 
and $J_{O_2}(z)$ represents the local specific rate of O$_2$ production and consumption in the reactor depth (in mol$_{O_2}$/kg$_X$/h). It is the result of photosynthetic production and respiration consumption, on which evolution depends on the local irradiance:
\begin{equation}
\label{Eq.JO2z}
J_{O_2}(z)=\rho_m \dfrac{K}{K+G(z)}\Phi'E_a G(z)-\dfrac{J_{NADH_2}}{\nu_{NADH_2-O_2}}\dfrac{K_R}{K_R+G(z)}
\end{equation}
where 
\begin{itemize}
\item $G(z)$ is the local irradiance, 
\item $\rho_m$ is the maximum value of the energetic yield, 
\item $E_a$ is the mass absorption coefficient, linked to the pigment content, 
\item $\Phi'$, $J_{NADH_2}$ and $\nu_{NADH_2-O_2}$ are stoichiometric yield expressed from the stochiometric equation of biomass synthesis, 
\item $K$ is the half saturation constant of photosynthesis, describing photosynthesis saturation with increasing light, 
\item $K_R$ is the respiration inhibition constant, describing the decrease of respiration in light. 
\end{itemize}
Most of the model parameters are constant, except $K$ and $K_R$ for which independent oxygen or fluorescence measurements are used for proposing values, which are strain dependent.

The two-flux model can then be used to model the irradiance $G(z)$ and the following formulation of irradiance distribution can be employed \cite{bib:Pottier}: 
\begin{equation}
\label{Eq.Gz}
G\left( z \right) = 2{q_0}\frac{{\left( {1 + \alpha } \right){e^{\delta \left( {L - z} \right)}} - \left( {1 - \alpha } \right){e^{ - \delta (L - z)}}}}{{{{\left( {1 + \alpha } \right)}^2}{e^{\delta L}} - {{\left( {1 - \alpha } \right)}^2}{e^{ - \delta L}}}}
\end{equation}
where $\delta  = X\sqrt {{E_a}\left( {{E_a} + 2b{E_s}} \right)} $ is the two-flux extinction coefficient, and 
$\alpha  = \sqrt {\left( {{E_a}} \right)/\left( {{E_a} + 2b{E_s}} \right)} $
the linear scattering modulus. $b$ is the backward scattering fraction (dimensionless). $E_s$, the mass scattering coefficients (m$^2$ kg$^{-1}$), and $E_a$ are chosen as a function of $q_0$, the incident light intensity. These optical properties are strain dependent and vary with growth conditions. Empirical formulae deduced from experiments in a wide range of incident light conditions \cite{bib:Takache2012} have been used here (see Table \ref{tab.modelecomplet}). 
 
\begin{table}[h]
	\caption{Model parameters}
	\label{tab.modelecomplet}
	\begin{center}
		\begin{tabular}{|l|c|c|}
			\hline
			Parameter & Value & Unit\\
			\hline
			$L$ & 0.05 & m\\
			\hline
			$b$ & 0.08& - \\
			\hline
			$E_a$ & $-28*\log(q_0) + 337$& m$^2$/kg\\
			\hline
			$E_s$ & $ 28.9*\log(q_0) + 708$ & m$^2$/kg\\
			\hline
			$K$ & 120 & $\mu$mol/m$^2$/s \\
			\hline	
			$\Phi'$ & 1.12 $10^{-7}$ & mol/$\mu$mol \\
			\hline
			$J_{NADH_2}/\nu_{NADH_2-O_2}$ & 3.19 $10^{-4}$& - \\
			\hline	
			$\nu_{O_2-X}$ & 1.183 & - \\
			\hline
			$M_x$& 24 $10^{-3}$& kg/C-mol \\
				\hline	
				$K_R$& 6 &$\mu$mol/m$^2$/s \\
			\hline		
			$\rho_m$& 0.8 & - \\
			\hline		
		\end{tabular}
	\end{center}
\end{table}

\section{Control strategies}
\label{sect.control}
\subsection{Problem formulation}

The aim is to get high cultivation in PBR, by improving the production and maintaining the quality of the product. One way to achieve this goal consists in regulating the biomass concentration at a given setpoint that leads to a high production of the PBR. In the case of continuous operation mode, at constant incident light, the biomass concentration can be controlled hydraulically through the dilution rate $D$ in open or closed-loop aiming maximum productivity, while avoiding washout, \textit{i.e.}, an unstable equilibrium point corresponding to the disappearance of the microorganisms from the cultivation system, where $X$=0. The photobioreactor can be operated at various concentrations of biomass in accordance with the selected working protocol. 
In this study, the regulation of the biomass concentration in a continuous PBR is considered, based on the so-called  \textit{turbidostat} protocol. The biomass concentration is assumed to be measured online via a turbidity sensor. It can also be determined from oxygen release measurements.

\subsection{Setpoint determination}
\begin{figure*}[t]
      \centering
       \includegraphics[scale = 0.44]{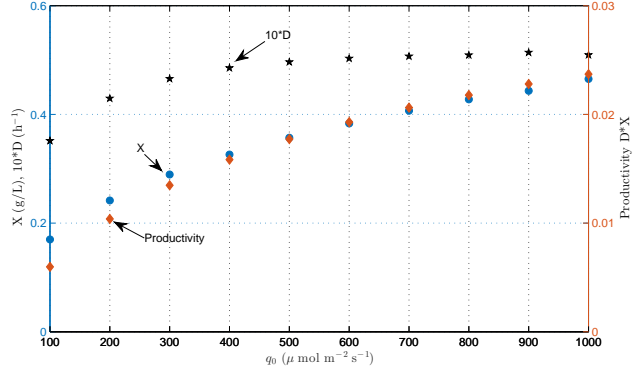}
      \caption{Setpoint of biomass concentration and corresponding productivity and dilution rate versus constant applied incident light intensity. }
      \label{Fig.Setpoints}
   \end{figure*}

\begin{figure}[thpb]
      \centering
       \includegraphics[scale = 0.47]{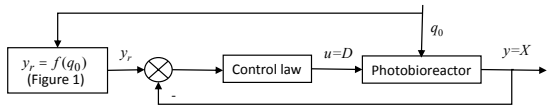}
      \caption{Block diagram of the control strategy. }
      \label{Fig.structure}
   \end{figure}

For constant incident light, the operating point (in terms of biomass concentration and dilution rate) can be determined so that the biomass productivity is maximized. The latter is defined as the product of steady state values of biomass concentration times dilution rate. It is equal to $X \times D$ at the equilibrium. 
Experimental protocol can be defined to determine the optimal setpoint as a function of the applied incident light intensity. Consequently, the result can be illustrated as shown by Fig. \ref{Fig.Setpoints} for $100 \leq q_0 \leq 1000$ $\mu$mol/m$^2$/s. It provides static values of $X$, $D$ and productivity $D \times X$ as functions of the incident light intensity $q_0$. Thus, for a given value of the incident light intensity, the reference value of the biomass concentration can be deduced from this figure.  
Hereafter, two control laws will be designed so that the system is operated at a given biomass concentration setpoint, that depends on the incident light intensity (deduced from Fig. \ref{Fig.Setpoints}). The control input is the dilution rate (or equivalently, the flow rate) and the output is the biomass concentration. The block diagram is depicted in Fig. \ref{Fig.structure}. 

\subsection{Model-based control}

System (\ref{Eq.dX}) is a single-input single-output (SISO) nonlinear control-affine model
\begin{equation}
\begin{array}{l}
\dot{X} = f_x(X) + f_u(X) u \\
y=h(X)=X
\end{array}
\label{eq.generale}
\end{equation} 
where $X$ is the biomass concentration, $u$ is the control input ($u=D$), $y$ is the output, $f_x$, $f_u$ and $h$ are nonlinear functions given by (\ref{Eq.dX}). An input-output feedback linearization is employed  \cite{kh}. 
With Equations \eqref{eq.generale}, where the state space is of dimension $1$, this linearization is of course equivalent to a static state feedback linearization. 
Classical linear controllers are thus employed. First, one must determine the \emph{relative degree} $r$. It is defined as the lowest order of the time derivative of $y$ that directly depends on the input $u$.
Here $r = 1$. 
Let $y_r$ be the reference trajectory. Suppose that the tracking error $e=y-y_r$ is specified by a first order dynamics as follows:
\begin{equation}
 \dot{e}+\lambda e=0
\label{eq.reference}
\end{equation} 
where the gain $\lambda>0$ is a tuning parameter. The linearizing feedback law is given by
\begin{equation}
 u=\left(-L_{f_x}h(X)-\lambda(y-y_r)\right)/\left(L_{f_u}h(X)\right)
\label{eq.CL_gen}
\end{equation} 
where $L_{f_{\bullet}}h(X)$ is the Lie derivative of $h$ with respect to the vector field $f_{\bullet}$.

Usually, this control law presents two main drawbacks:
\begin{itemize}
	\item its efficiency depends on the knowledge of the system dynamics,
	\item it assumes that all the state variables are available.
\end{itemize}

In our case, formulae (\ref{eq.CL_gen}) becomes
\begin{equation}
 D=\left(r_X+\lambda(y-y_r)\right)/X
\label{eq.CLin}
\end{equation} 
In order to be representative of the uncertainty of bioprocesses models, a simplified model of \eqref{Eq.rX} for the growth rate $r_X(X)$ is considered \cite{Bib:Cornet} in the control law (\ref{eq.CLin}): 
\begin{equation}
{r_X (X)} = \left[ {\mu_{0}}\frac{\overline{G}}{K_I + \overline{G} + {\overline{G}^2}/{K_{II}}} - \mu_r\right] X
\label{eq.mu_simple}
\end{equation} 
where $\mu_{0}$ is related to the maximal specific growth rate, $\mu_r$ represents the respiration rate, and $K_I$ (resp. $K_{II}$) is the limitation (resp. inhibition) constant \cite{bib:Dochain}. The mean incident light intensity $\overline{G}$ is given, with the same notations as previously, by:
\begin{equation}
\overline{G} = \frac{1}{L}\int_{0}^{L}{{q_0}{\rm{exp}}\left[ { - \frac{{\left( {1 + \hat{\alpha} } \right)}}{{2 \hat{\alpha} }}\hat{E_a} Xz} \right]dz}
\label{eq.G_moy}
\end{equation} 
The parameters of the simplified model were determined from the same data as those in \eqref{Eq.rX}-\eqref{Eq.Gz}. They are given by Table \ref{tab.SimplifiedModelParameters} \cite{bib:Fouchard}. 

\begin{table}[h]
\caption{Simplified model parameters}
\label{tab.SimplifiedModelParameters}
\begin{center}
\begin{tabular}{|l|c|c|c|c|c|c|}
\hline
Parameter & $\mu_0$ & $\mu_r$& $ \hat{\alpha}$ & $\hat{E_a}$ & $K_I$ & $K_{II}$\\
\hline
Value & 0.14& 0.013 &0.71 & 151 & 120 & 500\\
\hline
Unit & $h^{-1}$& $h^{-1}$ & - & m$^2$/kg & $\mu$mol/m$^{2}$/s & $\mu$mol/m$^{2}$/s\\	
\hline
\end{tabular}
\end{center}
\end{table}

\subsection{Model-free control and intelligent controllers\protect\footnote{See \cite{csm} for more details, and \cite{nice} for a hardware implementation.}}

\subsubsection{The ultra-local model}\label{ulm}
For simplicity's sake, let us restrict ourselves to SISO systems. The unknown global description of the plant is replaced by the \emph{ultra-local model}:
\begin{equation}
\dot{y} = \mathfrak{F} + \mathfrak{a} u \label{1}
\end{equation}
where:
\begin{itemize}
\item the control and output variables are $u$ and $y$,
\item the derivation order of $y$ is $1$ like in most concrete situations, 
\item $\mathfrak{a}  \in \mathbb{R}$ is chosen by the practitioner such that $\mathfrak{a} u$ and
$\dot{y}$ are of the same magnitude.
\end{itemize}
The following comments might be useful:
\begin{itemize}
\item Equation \eqref{1} is only valid during a short time lapse. It must be continuously updated,
\item $\mathfrak{F}$ is estimated via the knowledge of the control and output variables $u$ and $y$,
\item $\mathfrak{F}$ subsumes not only the unknown structure of the system, {which most of the time will be nonlinear}, but also of
any disturbance.
\end{itemize}

\subsubsection{Intelligent controllers}
Close the loop with the following \emph{intelligent proportional controller}, or \emph{iP},
\begin{equation}\label{ip}
u = - \frac{\mathfrak{F} - \dot{y}_r + K_P e}{\mathfrak{a}}
\end{equation}
where:
\begin{itemize}
\item $e = y - y_r$ is the tracking error,
\item $K_P$ is a usual tuning gain.
\end{itemize}
Combining \eqref{1} and \eqref{ip} yields:
\begin{equation}\label{ref}
\dot{e} + K_P e = 0
\end{equation}
where $\mathfrak{F}$ does not appear anymore. The tuning of $K_P$ is therefore quite straightforward. This is a major benefit when
compared to the tuning of ``classic'' PIDs (see, \emph{e.g.},
\cite{astrom}, and the references therein). Some more comments may be useful:
\begin{itemize}
\item See \cite{ecc14} for a connection with stability margins.
\item Equations \eqref{ip} and \eqref{ref} render pointless any formal checking of the closed-loop system behavior (compare, \textit{e.g.}, with \cite{donze}).
\end{itemize}

\subsubsection{Estimation of $\mathfrak{F}$}\label{F}
Assume that $\mathfrak{F}$ in \eqref{1} is ``well'' approximated by a piecewise constant function ${\mathfrak{F}}_{\text{est}} $.
\begin{enumerate}
\item Rewrite \eqref{1}  by using the well-known notations from operational calculus: 
$$
sY = \frac{\Phi}{s} + \mathfrak{a} U +y(0)
$$
where $\Phi$ is a constant. We get rid of the initial condition $y(0)$ by multiplying both sides on the left by $\frac{d}{ds}$:
$$
Y + s\frac{dY}{ds}=-\frac{\Phi}{s^2} + \mathfrak{a} \frac{dU}{ds}
$$
Noise attenuation is achieved by multiplying both sides on the left by $s^{-2}$. It yields in the time domain the realtime estimate, thanks to the equivalence between $\frac{d}{ds}$ and the multiplication by $-t$,
\begin{equation*}\label{integral}
{\small {\mathfrak{F}}_{\text{est}}(t)  =-\frac{6}{\tau^3}\int_{t-\tau}^t \left\lbrack (\tau -2\sigma)y(\sigma) + \mathfrak{a}\sigma(\tau -\sigma)u(\sigma) \right\rbrack d\sigma }
\end{equation*}
where $\tau > 0$ might be quite small. This integral, which is a low pass filter, may of course be replaced in practice by a classic digital filter.
\item Close the loop with the iP \eqref{ip}. It yields:
$$
{\mathfrak{F}}_{\text{est}}(t) = \frac{1}{\tau}\left[\int_{t - \tau}^{t}\left(\dot{y}_r - \mathfrak{a} u
- K_P e \right) d\sigma \right] 
$$
\end{enumerate}

\section{Simulation tests}
\label{sect.simulation}
\subsection{Setpoint change with light}
\label{cas1}
\begin{figure*}[thpb]
      \centering
       \includegraphics[scale=0.495]{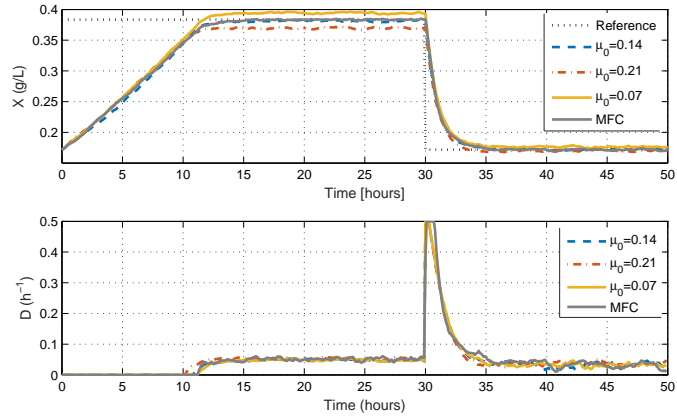}
      \caption{Comparison of closed-loop responses to a setpoint change (light decreasing step change).}
      \label{Fig-C1}
   \end{figure*}

First, the two control laws are compared in the case of a piecewise-constant light intensity with the following profile: 
\begin{eqnarray}
\label{eq:modeleq0}
\left \{
\begin{array}{cc}
q_0 (t)=600 \;\mu \text{ mol m}^{-2}\text{ s}^{-1} & \text{for } 0 < t \leq 30 h \\
q_0 (t) =100 \;\mu \text{ mol m}^{-2}\text{ s}^{-1}& \text{otherwise.} 
\end{array}\right.
\end{eqnarray} 
>From \eqref{eq:modeleq0} and Fig. \ref{Fig.Setpoints}, the biomass concentration reference value is deduced. It is piecewise constant: it varies from 0.17 to 0.38 kg/m$^3$ at time $t=30$ hours.
The biomass concentration at initial time is $X$(0)=0.17 kg/m$^3$, and the simulation duration is set to $50$ hours. The output is assumed to be measured with a sampling time $T_s = 6$ min and to be corrupted by an additive zero-mean white Gaussian noise with a standard deviation of about $1$\%. 
The model \eqref{Eq.rX}-\eqref{Eq.Gz} is used for the plant, whereas the simplified model  \eqref{eq.mu_simple}-\eqref{eq.G_moy} is considered for the model-based controller \eqref{eq.CLin}.
The tuning parameters of the control law are as follows: $\lambda = 1$ for the model-based law and $(\mathfrak{a}, K_p, \tau) = (0.2, 5, 15 T_s)$ for the model-free one. The control laws are implemented using a zero-order hold.
The control input $D$ satisfies the constraints  $0~\leq~D~\leq~0.5~h^{-1}$.
For the robustness study, only the uncertainty on the variable $\mu_0$ is considered, since the latter is the most influential parameter. 
Three cases are considered: $\mu_0$=0.14 (nominal value), $\mu_0$=0.21 and $\mu_0$=0.07~h$^{-1}$. These values are chosen from the confidence interval on this variable \cite{bib:Fouchard}.
The results are depicted in Fig. \ref{Fig-C1}.

The two control laws achieve the tracking of the reference biomass concentration, with similar time responses. First, the control is canceled, {\it i.e.}, the system operates in the batch mode, so that the biomass concentration reaches almost the reference value. Then, after about $10$ hours, the control input switches to a continuous mode and a dilution is applied so that the output is maintained at the desired value. At time $t=30$ hours onwards, the reference value decreases to 0.17~kg/m$^3$. Consequently, the controllers apply a higher dilution rate to dilute the culture and attain the new reference value (after about 5 h). Then, the dilution rate reaches a constant value that maintains the output at this new reference. It can be noticed that the model-based controller is sensitive to the value of the parameter $\mu_0$. Indeed, the reference tracking presents an offset in this case. The response with the model-free controller on the other hand is offset-free. Its robustness is highlighted.
   

\subsection{Light change with constant setpoint}
\label{cas2}
\begin{figure}[thpb]
      \centering
       \includegraphics[scale=.42]{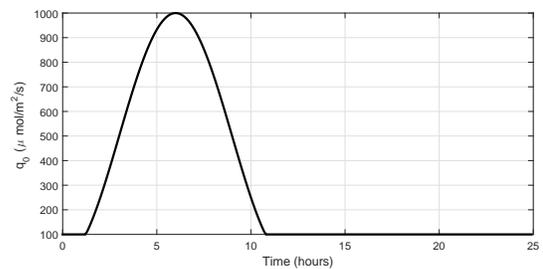}
      \caption{Incident light evolution versus time. }
      \label{Fig-C2p}
   \end{figure}

\begin{figure*}[thpb]
     \centering
       \includegraphics[scale=0.495]{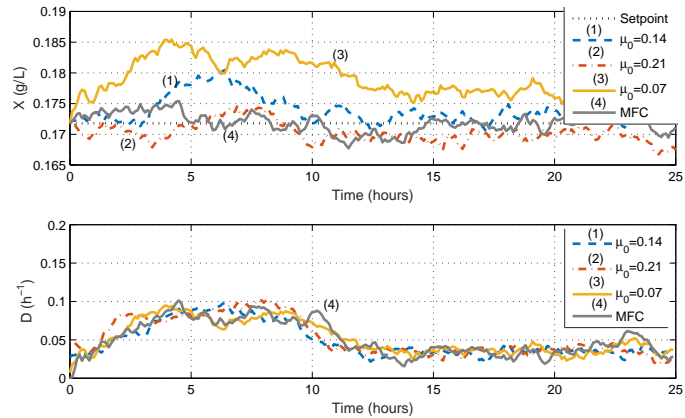}
      \caption{Comparison of closed-loop responses to a constant setpoint (time-varying incident light).}
      \label{Fig-C2}
   \end{figure*}

The controllers are now compared in the case of a constant biomass concentration setpoint $y_r = 0.175$ kg/m$^3$, with a time-varying incident light intensity. The latter is assumed to follow a profile depicted in Fig. \ref{Fig-C2p}, \textit{i.e.}, 
it is chosen in order to be similar to a day/night cycle of a solar light except the minimum level of incident light, which has been set to 100 $\mu$mol/m$^2$/s, here. 
The controllers tuning parameters and the simulation conditions are similar to those considered in Section \ref{cas1}. 
Simulation results, illustrated by Fig. \ref{Fig-C2}, show that despite the light variation, considered here as a disturbance, the two control laws maintain the output at its reference value. Indeed, the dilution rate is modified, according to the incident light profile. Nevertheless, the model-free controller performs better than the model-based one. The model-based controller is again sensitive to the uncertainty of $\mu_0$. Its drawback is here also underlined.

%

\section{Conclusion}
\label{sect.conclusion}
The model-free control strategy yields better performances than the model-based one,\footnote{Those comparisons will be more closely investigated elsewhere.} in terms of 
\begin{itemize}
\item setpoint tracking and robustness with respect to model uncertainties,
\item no modeling perequisite,  \textit{i.e.},
\begin{itemize}
\item only the output variable needs to be known,
\item the estimation of the state variables becomes useless,
\end{itemize}
\item easy and costless implementation.
\end{itemize}
An experimental set-up and the control of the specific growth rate for insuring a high level of biomass growth are now being developed. They will hopefully
be presented soon. The above preliminary results should not only be confirmed but also amplified.


\addtolength{\textheight}{-0,7cm}   




\begin{thebibliography}{99}


\bibitem{kim}
S.-K. Kim (Ed.), Handbook of Marine Microalgae: Biotechnology Advances, Academic Press, 2015.
\bibitem{spo}
P. Spolaore, C. Joannis-Cassan, E. Duran and A. Isambert, Commercial applications of microalgae, J. Biosci. Bioengin., vol. 101, pp. 87--96, 2006.

\bibitem{bib:Chevalier} 
P. Chevalier, D. Proulx, P. Lessard, W.F. Vincent and J. de la Noue, Nitrogen and phosphorus removal by high latitude mat-forming cyanobacteria for potential use in tertiary wastewater treatment, J. Appl. Phycol., vol. 12, pp. 105--112, 2000.


\bibitem{bib:Chisti} Y. Chisti, Biodiesel from microalgae, Biotechno. Adv.,  vol. 25, pp. 294--306, 2007.

\bibitem{goh}
C.S. Goh and K.T. Lee,  A visionary and conceptual macroalgae-based third-generation bioethanol (TGB) biorefinery in Sabah, Malaysia as an underlay for renewable, and sustainable development, Renew. Sustain. Energ. Rev., vol. 14, pp. 842--848, 2010.


\bibitem{bib:pruvost} J. Pruvost, Cultivation of algae in photobioreators for biodiesel production, in Biofuels:alternative feedstocks and conversion processes, Pandley et al. Elsevier, 2011.


\bibitem{bib:Jiang}  L. Jiang, S. Luo, X. Fan, Z. Yang and R.  Guo, Biomass and lipid production of marine microalgae using municipal wastewater and high concentration of CO$_2$, Applied Energ., vol. 88, pp. 3336--3341, 2011.

\bibitem{bib:Lam} M.K. Lam, K.T. Lee and A.R. Mohamed, Current status and challenges on microalgae-based carbon capture, Int. J. Greenhouse Gas Contr., vol. 10, pp. 456--469, 2012.

\bibitem{bib:Pandey} A. Pandey, Biofuels: Alternative Feedstocks and Conversion Processes. Academic Press, 2011.

\bibitem{razzak}
S.A. Razzak, M.M. Hossain, R.A. Lucky, A.S. Bassi and H. de Lasa,  Integrated CO$_2$ capture, wastewater treatment and biofuel production by microalgae culturing -- A review, Renew. Sustain. Energ. Rev., vol. 27, pp. 622--653, 2013.

\bibitem{iste}
S. Tebbani, F. Lopes, R. Filali, D. Dumur and D. Pareau, CO$_2$ Biofixation by Microalgae: Modeling, estimation and control. ISTE -- Wiley, 2014.


\bibitem{bib:Abdollahi} J. Abdollahi and S. Dubljevic, Lipid production optimization and optimal control of heterotrophic microalgae fed-batch bioreactor, Chem. Eng. Sci., vol. 84, pp. 619--627, 2012.

\bibitem{and}
G.A. de Andrade, M. Berenguel, J. L. Guzm\'{a}n, D.J. Pagano,
F.G. Aci\'{e}n, Optimization of biomass production in outdoor
tubular photobioreactors, J. Proc. Contr., vol. 37,
pp. 58--69, 2016.

\bibitem{pde}
I. Fern\'{a}ndez, F.G. Aci\'{e}n, M. Berenguel, J.L. Guzm\'{a}n, First principles model of a tubular photobioreactor for microalgal production, 
Ind. Eng. Chem. Res., vol. 53, pp. 11121--11136, 2014.

\bibitem{ifrim}
G.A. Ifrim, M. Titica, M. Barbu, L. Boillereaux, G. Cogne, S. Caraman and J. Legrand, Multivariable feedback linearizing control of \textit{Chlamydomonas reinhardtii} photoautotrophic growth process in a torus photobioreactor, Chem. Engin. J., vol. 218, pp. 191--203, 2013.  


\bibitem{mail}
L. Mailleret, O. Bernard and J.P. Steyer, Nonlinear adaptive control for bioreactors with unknown kinetics, Automatica, vol. 40, pp. 1379--1385, 2004.

\bibitem{rod}
A. E. Rodr\'{\i}guez, R. Luna, J. R. P\'{e}rez, J. Torres, A. Dom\'{\i}nguez, H. Sira and R. Castro, Robust control for cultivation of microorganisms in a high density fed-batch bioreactor (in Spanish), IEEE Latin Amer. Trans., vol. 13, pp. 1927--1933, 2015.

\bibitem{bib:tebbani2015} S. Tebbani, F. Lopes and G. Becerra Celis,  Nonlinear control of continuous cultures of Porphyridium purpureum in a photobioreactor, Chem. Eng. Sci., vol. 123, pp. 207--219, 2015.


\bibitem{paw}
A. Pawlowski, J.L. Mendoza, J.L. Guzm\'{a}n, M. Berenguel, F.G. Aci\'{e}n, S. Dormido, 
Effective utilization of flue gases in raceway reactor with event-based pH control for microalgae culture, Bioresource Techno., vol. 170, pp. 1--9, 2014.

\bibitem{bib:Dochain} D. Dochain (Ed.), Automatic Control of Bioprocesses. ISTE -- Wiley, 2008.



\bibitem{bib:Bernard} O. Bernard, Hurdles and challenges for modelling and control of microalgae for CO$_2$ mitigation and biofuel production, J. Process Contr., vol. 21, pp. 1378--1389, 2011.

\bibitem{csm}
M. Fliess and C. Join, Model-free control, Int. J. Contr., vol. 86, pp. 2228--2252, 2013.

\bibitem{nice}
C. Join, F. Chaxel and M. Fliess, ``Intelligent'' controllers on cheap and small programmable devices,
2nd Int. Conf. Contr. Fault-Tolerant Syst., Nice, 2013. Online: \newline
{\tt \scriptsize https://hal.archives-ouvertes.fr/hal-00845795/en/}



\bibitem{toulon}
F. Lafont, J.-F. Balmat, N. Pessel and M. Fliess,
A model-free control strategy for an experimental greenhouse with an application to fault accommodation, Comput. Electron. Agricult., vol. 110, pp. 139--149, 2015.

\bibitem{berlin}
T. MohammadRidha, C.H. Moog, Model free control for type-1 diabetes: A fasting-phase study, 9th IFAC Symp. Biol. Medic. Syst., Berlin, 2015.

\bibitem{siam}
T. MohammadRidha, C.H. Moog, E. Delaleau, M. Fliess and C. Join, A variable reference trajectory for model-free glycemia regulation, SIAM Conf. Control Appl., Paris, 2015. Online: \newline
{\tt \scriptsize https://hal.archives-ouvertes.fr/hal-01141268/en/}

\bibitem{bara}
O. Bara, M. Fliess, C. Join, J. Day and S.M. Djouadi, Model-free immune therapy: A control approach to acute inflammation, Europ. Contr. Conf., Aalborg, 2016. Online: \newline
{\tt \scriptsize https://hal.archives-ouvertes.fr}



\bibitem{bib:Ifrim2014} G. A. Ifrim, M. Titica , G. Cogne, L. Boillereaux, J. Legrand and S. Caraman, Dynamic pH model for autotrophic growth of microalgae in photobioreactor: A tool for monitoring and control purposes. AIChE J., vol. 60, pp. 585-599, 2014.


\bibitem{bib:Takache2012} H. Takache, J.F. Cornet and J. Pruvost, Kinetic modeling of the photosynthetic of  \textit{chlamydomonas reinhardtii} in a photobioreactor, Biotechno. Progr., vol. 28, pp. 681--692, 2012.


\bibitem{bib:Pottier} L. Pottier, J. Pruvost, J. Deremetz, J.F Cornet, J. Legrand and J.G.~Dussap, A fully predictive model for one-dimensional light attenuation by {\it Chlamydomonas reinhardtii} in a torus reactor, Biotechnol. Bioengin., vol. 91, pp. 569--582, 2005.


\bibitem{kh}
H.K. Khalil, Nonlinear Systems (3rd ed.), Prentice Hall, 2002.

\bibitem{Bib:Cornet}
J.F. Cornet and C.G. Dussap, A simple and reliable formula for assessment of maximum volumetric productivities in photobioreactors, Biotechno. Prog., vol. 25, pp. 424--435, 2009.


\bibitem{bib:Fouchard}	S. Fouchard, J. Pruvost, B.  Degrenne,  M. Titica and J. Legrand, Kinetic modeling of light limitation and sulfur deprivation effects in the induction of hydrogen production with {\it Chlamydomonas reinhardtii}: Part I. Model development and parameter identification, Biotechno. Bioengin., vol. 102, pp. 232--245, 2009.

\bibitem{astrom}
K.J. {\AA}str\"om and T. H\"agglund, Advanced PID
Control. Instrument Soc. Amer., 2006.

\bibitem{ecc14}
M. Fliess and C. Join, Stability margins and model-free control: A first look, Europ. Contr. Conf., Strasbourg, 2014. Online: \newline
{\tt \scriptsize https://hal.archives-ouvertes.fr/hal-00966107/en/}


\bibitem{donze}
X. Jin, A. Donz\'{e}, J.V. Deshmukh and S.A. Seshia, Mining requirements from closed-loop control models, IEEE Trans. Comput.-Aided Design Integr. Circ. Syst., vol. 34, pp. 1704--1716, 2015.

\end{thebibliography}
\end{document}